\documentclass[superscriptaddress,nofootinbib
]{revtex4-2}

\usepackage{graphicx}
\usepackage{dcolumn}
\usepackage{bm}
\usepackage{amsmath} 
\usepackage[table,svgnames,dvipsnames]{xcolor}
\usepackage[mathlines]{lineno}
\usepackage[normalem]{ulem} 
\usepackage{aas_macros}
\usepackage{subfigure}
\usepackage{graphicx}
\usepackage{graphics}
\usepackage{dcolumn}
\usepackage{bm}
\usepackage{cases}
\usepackage{booktabs}
\usepackage{comment}
\usepackage{multirow}
\usepackage{makecell}
\usepackage{siunitx}
\usepackage{tabularx}
\usepackage{xspace}
\usepackage{soul} 
\graphicspath{{figs/}}
\usepackage{fontawesome}
\usepackage{hyperref} 

\usepackage{orcidlink} 
\usepackage{hanging} 

\renewcommand{\arraystretch}{1.4}
\def\be{\begin{equation}}
\def\ee{\end{equation}}

\def\ba#1\ea{\begin{align*}#1\end{align*}}

\renewcommand{\emph}[1]{\textit{#1}}
\definecolor{RoyalBlue}{rgb}{0.25,.41,.88}
\definecolor{WildStrawberry}{HTML}{EE2967}
\definecolor{RedWine}{rgb}{0.743,0,0}
\definecolor{bittersweet}{rgb}{1.0, 0.44, 0.37}
\definecolor{burntorange}{rgb}{0.8, 0.33, 0.0}
\definecolor{midnightgreen}{rgb}{0.0, 0.29, 0.33}
\definecolor{otherblue}{rgb}{0.20, 0.73, 0.92}
\definecolor{purple}{rgb}{0.647,0.439,0.882}
\definecolor{UltraViolet}{HTML}{6433FF}

\usepackage[nameinlink,noabbrev]{cleveref}
\crefname{equation}{Eq.}{Eqs.}
\crefname{section}{Section}{Sections}
\crefname{figure}{Figure}{Figures}
\crefname{table}{Table}{Tables}
\crefname{appendix}{Appendix}{Appendices}
\Crefname{figure}{Figure}{Figures}
\Crefname{equation}{Equation}{Equations}
\Crefname{section}{Section}{Sections}
\Crefname{table}{Table}{Tables}

\newcommand{\mksym}[1]{\ifmmode {\rm #1}\else #1\fi}

\newcommand{\Om}{\Omega_\mathrm{m}}

\newcommand{\lcdm}{$\Lambda$CDM}

\newcommand{\rvev}{\rho_{\rm VEV}}  
\newcommand{\fvev}{f_{\rm VEV}}  
\newcommand{\wde}{w_{\rm de}}  
\newcommand{\rde}{\rho_{\rm de}}  
\newcommand{\wowa}{w_0w_a} 
\newcommand{\rhol}{\rho_\Lambda}

\begin{document}

\title{
Vacuum Phase Transition and Decay to Cross the Phantom Divide
} 

\author{William Giar\`e\orcidlink{0000-0002-4012-9285}} 
\email{giare@hawaii.edu} 
\affiliation{Department of Physics and Astronomy, University of Hawai‘i, Honolulu, HI 96822, USA} 

\author{Kushal Lodha\orcidlink{0009-0004-2558-5655}}
\email{klodha@bu.edu}
\affiliation{Department of Physics, Boston University, Boston, MA 02215, USA}
\affiliation{Korea Astronomy and Space Science Institute, 776, Daedeokdae-ro, Yuseong-gu, Daejeon 34055, Republic of Korea}
\affiliation{University of Science and Technology, 217 Gajeong-ro, Yuseong-gu, Daejeon 34113, Republic of Korea}

\author{Robert R.\ Caldwell\orcidlink{0000-0001-7490-7463}}
\email{robert.r.caldwell@dartmouth.edu} 
\affiliation{Department of Physics and Astronomy,
Dartmouth College, Hanover, NH 03755, USA} 

\author{Eleonora Di Valentino\orcidlink{0000-0001-8408-6961}}
\email{e.divalentino@sheffield.ac.uk}
\affiliation{School of Mathematical and Physical Sciences, University of Sheffield, Hounsfield Road, Sheffield S3 7RH, United Kingdom}

\author{Eric V.\ Linder\orcidlink{0000-0001-5536-9241}}
\email{evlinder@lbl.gov}
\affiliation{Berkeley Center for Cosmological Physics \& Berkeley Lab,
University of California, Berkeley, CA 94720, USA}

\begin{abstract}

We show that the vacuum metamorphosis model with decay fits current baryon acoustic oscillation and supernova distance data and cosmic microwave background measurements significantly better than $\Lambda$CDM and nearly as well as $w_0w_a$CDM with the same number of parameters. 
Vacuum metamorphosis with decay provides a physically motivated model of dark energy and an explanation of current cosmological data. 
In this scenario, non-perturbative quantum effects of a massive scalar field freeze the Ricci scalar curvature when it reaches a critical value proportional to the mass squared, inducing a phase transition in the scalar field, giving a phantom dark energy equation of state $w<-1$. After the phase transition, vacuum energy decays into light particles, providing a natural mechanism for the dark-energy equation of state to cross the phantom divide from $w<-1$ to $w>-1$, in agreement with current cosmological indications, without requiring an explicit coupling to matter. 

\end{abstract}

\maketitle

\section{Introduction}

Current cosmological distance data point toward 
a beautifully bizarre picture of cosmic acceleration, 
where the acceleration is stronger than expected from the \lcdm\ 
model of a cosmological constant plus cold dark matter at 
redshifts $z \gtrsim 0.5$, 
and somewhat weaker 
at $z \lesssim 0.5$ \cite{DESI:2025zgx,DESI:2025fii}. 

In terms of the dark energy equation of state, this 
indicates an evolution from $w < -1$ to $w > -1$, 
a so-called crossing of the phantom divide. 
Any two of the cosmological probes of 
baryon acoustic oscillations (BAO), cosmic microwave background (CMB), and Type Ia supernovae 
(SN) combine to give $\sim3\sigma$ evidence for phantom crossing (e.g.\ evaluated in a model-independent manner using Gaussian processes), 
while all three combine to give a $\sim 3.2$--$3.4\sigma$ indication against a cosmological constant within $w_0w_a$CDM \cite{DESI:2025fii,DES:2025sig,Hoyt:2026fve}. 

The physics behind crossing the phantom divide must be 
quite special, violating assumptions usually made about 
dark energy~\cite{Vikman:2004dc,Kunz:2006wc}; moreover, the paucity of dark energy 
density (equivalently $w$ well below $-1$) at early times 
and the required rapidity of the crossing add additional hurdles~\cite{Linder:2024rdj}. Nonetheless, the evidence for a 
phantom-divide crossing is robust across current data 
combinations~\cite{DESI:2025fii,Ozulker:2025ehg,Giare:2026oti}. Mechanisms that can potentially 
enable the required conditions include interaction of dark energy 
with either a matter sector~\cite{Das:2005yj} or another dark energy field~\cite{Zhao:2006mp}, 
modified gravity~\cite{Carroll:2004hc}, or a phase transition~\cite{Caldwell:2005xb}. 
However, coupling dark 
energy to matter, or nonminimal coupling to gravity, 
may induce problems with growth-of-structure data, 
CMB lensing, or the integrated Sachs--Wolfe effect~\cite{Chudaykin:2024gol,Wolf:2024stt,Ishak:2024jhs,Linder:2025zxb,Ye:2026yqk,Desmond:2026lds} 
(some specific models appear in~\cite{Dutta:2010kb,Giare:2024smz,Khoury:2025txd,Wolf:2025jed,Chen:2025ywv,Goh:2025upc,Liu:2025bss,Caldwell:2025inn,Li:2026xaz}, 
for example). 
Multiple interacting dark energy fields 
still require one of the fields to violate the canonical 
assumptions (i.e.\ $\rho>0$ and $w>-1$; see e.g.\ \cite{Linder:2025zxb}). 

Here we consider the phase-transition alternative in the 
form of the physically well-founded vacuum metamorphosis 
theory of Parker and collaborators~\cite{Parker:1999fc,Parker:2001ws,Parker:2003as,Caldwell:2005xb}, 
previously analysed in Refs.~\cite{DiValentino:2017rcr,DiValentino:2020kha} 
as a possible solution to the Hubble tension~\cite{DiValentino:2021izs,Abdalla:2022yfr,CosmoVerse:2025txj,H0DN:2025lyy}. 
This framework naturally generates phantom dark energy but requires one additional physical ingredient 
to cross the phantom divide: not a coupling to standard matter, but a 
decay of the vacuum energy into lighter particles, a 
feature that is not uncommon in fundamental physics.

In \Cref{sec:implementation} we describe the phantom vacuum metamorphosis theory, 
the effects of the phase transition, and the evolution afterward. \Cref{sec:datacode} discusses the code implementing 
the cosmological equations and the data sets used to constrain 
the theory. \Cref{sec:results} presents the results for fitting 
the parameters and compares the goodness of fit against the 
$w_0w_a$CDM and \lcdm\ models. We summarize and conclude in 
\Cref{sec:concl}.

\section{Vacuum Metamorphosis Framework}\label{sec:implementation}

The Ricci scalar $R$ plays a central role in cosmology.
It enters the Einstein--Hilbert action and, in a
homogeneous and isotropic universe, determines the
kinematics,
\be
R = 6\left(\dot H + 2H^2 + k a^{-2}\right)
  = 6H^2(1-q) + 6k a^{-2}\ , \label{eq:rq}
\ee
where $H=\dot a/a$ is the Hubble expansion parameter
and $q=- a\ddot a/\dot a^2$ is the deceleration
parameter. Henceforth we will take a flat universe, $k=0$.

Quantum cosmology effects often add higher powers of
$R$ to the action, as in Starobinsky $R^2$ gravity~\cite{Starobinsky:1980te},
Sakharov's induced gravity~\cite{Sakharov:1967pk},
and \"ubergravity~\cite{Khosravi:2017hfi}. These are
perturbative attempts to include quantum corrections.
Parker and collaborators
\cite{Parker:1999fc,Parker:2001ws,Parker:2003as,Caldwell:2005xb}
succeeded in obtaining a nonperturbative result for
the quantum effects of a massive (but light) scalar
field. There, the Ricci scalar acts as an order parameter
for a gravitational phase transition, occurring when
$R$ decreases to a critical value,
\begin{equation}
    R = \chi m^2 \,,
\end{equation}
where $m$ is the mass of the scalar field and $\chi$ is
a constant of order unity.

Throughout most of cosmic history, the curvature is
``supercritical'', $R > \chi m^2$, and the expansion
history is governed by the usual matter and radiation
components, plus any scalar field vacuum expectation value
(VEV), as in the standard Friedmann equations
(where the VEV acts like a cosmological constant).
Once $R$ reaches the critical value, a non-perturbative
feedback mechanism, often described as a gravitational
analogue of Lenz's law~\cite{Parker:2003as}, acts to
prevent $R$ from decreasing further, effectively freezing
the Ricci scalar at $R = \chi m^2$ on cosmological scales.
From \cref{eq:rq}, we immediately see that this leads
to cosmic acceleration (despite the loss of the
false vacuum VEV in the phase transition): as the universe
ages and $H$ decreases, $1-q$ must increase to preserve
$R = \chi m^2$, meaning that $q$ must eventually
become negative (alternately, as the matter energy density $\rho_m$ dilutes, the effective dark energy density $\rde$ must increase to preserve constant $R$, indicating a phantom behavior).

Without even solving the equations of motion, we can understand
how the expansion rate behaves after the phase transition.
First, any constant vacuum energy simply contributes a constant
to $H^2$ and to $R$.
Next, recall that the radiation stress-energy tensor is traceless
and does not contribute to $R$, so the post-transition
vacuum energy can have a component that acts like radiation.
Matter energy density, however, does make $R$ evolve, so we must
cancel the contribution that standard matter density brings to
$R$ with a new ``vacuum matter'' component with a negative energy
density.
We emphasize that this in no way affects standard matter,
which remains present and does not interact with any of the
three vacuum energy components, which we will call vacuum
$\lambda$ (or $v\lambda$), vacuum radiation (or $vr$), and vacuum
matter (or $vm$). In summary, the vacuum energy density is
(cf.\ \cite{Parker:2003as})
\be
\rho_{\rm de}=
\begin{cases}
\rvev\ ,\quad z>z_\ast\\
\rho_{v\lambda}+\rho_{vr}+\rho_{vm}\ ,\quad z\le z_\ast\ ,
\end{cases} \label{eq:rhodetrs}
\ee
where $\rvev$ is the pre-transition false VEV.

While the vacuum phase transition
gives a phantom equation of state, $w_{\rm de} < -1$,
just after the transition at $z_\ast$ (i.e.\ when $R = \chi m^2$),
in the original model the later behavior rapidly approaches
$w_{\rm de} = -1$ and does not cross $-1$
\cite{DiValentino:2017rcr,DiValentino:2020kha}.
Coupling such a phantom model to the matter sector could
uplift it through $w = -1$
(e.g.\ see~\cite{Linder:2025zxb}), but this is
arbitrary and causes issues with cosmic structure growth,
as mentioned above. However, it is quite natural in many
physics models for the scalar field to decay into light particles,
effectively scalar radiation~\cite{Parker:2003as}. Indeed, one generally needs to
impose some symmetry to prevent such decays. This decay
serves to uplift the equation of state and enables it to
cross the phantom divide. We can write the decay
heuristically, with exact equations given below, as
\begin{equation}
    \{v\lambda, vm, vr\} \rightarrow vd \ ,
\end{equation}
where $vd$ denotes the decay products (light particles) of
the scalar field.
We emphasize that there is a single scalar field, and it is
merely for convenience that we talk about it in terms of
three vacuum components.
This is the new vacuum metamorphosis with decay 
model (abbreviated simply as VM), and it updates Eq.~(\ref{eq:rhodetrs}) to
\be
\rho_{\rm de}=
\begin{cases}
\rvev\ ,\quad z>z_\ast\\
\rho_{v\lambda}+\rho_{vr}+\rho_{vm}+\rho_{vd}\ ,
\quad z\le z_\ast\ .
\end{cases} \label{eq:rhodedecay}
\ee

To recap, before the phase transition the total energy density
entering the Friedmann equations for the expansion is
\be
\rho_{\rm tot}(z) = \rho_m(z) + \rho_r(z)+\rvev\ ,
\quad z>z_\ast\ ,
\ee
consisting of matter, radiation, and a constant false VEV.
After the phase transition,
\be
\rho_{\rm tot}(z) = \rho_m(z) + \rho_r(z) + \rho_{\rm de}(z)\ ,
\quad z<z_\ast\ ,
\ee
with effective dark energy density and equation of state
\begin{align}
    \rho_{\rm de}(z) &=
    \rho_{v\lambda}(z) + \rho_{vm}(z) +
    \rho_{vr}(z) + \rho_{vd}(z)\ , \label{eq:rhov}\\
    w_{\rm de}(z) &=
    \frac{-\rho_{v\lambda}(z) + \rho_{vr}(z)/3 + \rho_{vd}(z)/3}
    {\rho_{v\lambda}(z) + \rho_{vm}(z) + \rho_{vr}(z) + \rho_{vd}(z)}\ .
\end{align}

For the form of the decay, we take the standard
radiative decay into light particles,
\begin{align}
    \frac{\mathrm{d}\rho_{v\lambda}}{\mathrm{d}N}
    &= -\frac{\Gamma}{H}\,\rho_{v\lambda}\,,
    \label{eq:vm_rhovl_evol}\\
    \frac{\mathrm{d}\rho_{vm}}{\mathrm{d}N}
    &= -3\rho_{vm} - \frac{\Gamma}{H}\,\rho_{vm}\,,
    \label{eq:vm_rhovm_evol}\\
    \frac{\mathrm{d}\rho_{vr}}{\mathrm{d}N}
    &= -4\rho_{vr} - \frac{\Gamma}{H}\,\rho_{vr}\,,
    \label{eq:vm_rhovr_evol}\\
    \frac{\mathrm{d}\rho_{vd}}{\mathrm{d}N}
    &= -4\rho_{vd} + \frac{\Gamma}{H}
    \left(\rho_{v\lambda} + \rho_{vm}+\rho_{vr}\right)\ ,
    \label{eq:vm_rhovd_evol}
\end{align}
where $N \equiv -\ln(1+z)$ is the expansion e-fold
parameter, and $\Gamma$ is the constant decay rate
(hence the decay is exponential in time).
Note that $\Gamma$ does not enter explicitly into the
equation of state $\wde$ since the decay is an exchange
among dark energy components, although the decay does
impact the specific evolution of each component $\rho_{vi}$.
One might wonder why we don't combine the $vr$ and $vd$
components into a single radiation-like component; the answer
is that their perturbative sectors differ, as we discuss
in Appendix~\ref{sec:apxpert}.

Finally, we need the junction conditions at the phase transition. Matching 
$H$ and $\dot H$ (or, equivalently, $q$) across the
transition, one has
\begin{align}
    \rho_{v\lambda}(z_\ast) &=
    \rvev+\tfrac{1}{4}\,\rho_m(z_\ast)\ ,
    \label{eq:vli}\\
    \rho_{vm}(z_\ast) &= -\rho_m(z_\ast)\ ,
    \label{eq:vmi}\\
    \rho_{vr}(z_\ast) &= \tfrac{3}{4}\,\rho_m(z_\ast)\ ,
    \label{eq:vri}\\
    \rho_{vd}(z_\ast) &= 0\ .
    \label{eq:vdi}
\end{align}
Note that
\begin{equation}
    \rho_{v\lambda}(z_\ast) + \rho_{vm}(z_\ast) +
    \rho_{vr}(z_\ast) + \rho_{vd}(z_\ast)= \rvev\ ,
    \label{eq:vm_rhovac_sum}
\end{equation}
as \cref{eq:rhodetrs} requires. 

We parametrize the decay by the dimensionless quantity
$\gamma_\ast\equiv\Gamma/H(z_\ast)$, expected to be of order
unity. If the universe is approximately matter dominated
at the transition, $m^2\approx 3H^2(z_\ast)$, so
$\gamma_\ast\sim1$ corresponds to a decay rate comparable
to the scalar-field mass. 
We characterize the false-vacuum energy density through
the ratio $\rvev/\rho_\Lambda$, where $\rho_\Lambda$ is the vacuum energy density of the flat \lcdm\
reference model with the same $H_0$ and standard-matter
density parameter $\Omega_m$ as the VM model.
Here $\Omega_m$ includes only standard matter (baryons,
cold dark matter, and massive neutrinos), excluding all
components of the VM sector. Since $\rho_\Lambda$ does not explicitly depend on $z_\ast$, the ratio $\rvev/\rho_\Lambda$ directly indicates whether
the vacuum energy density before the phase transition
is lower than, higher than, or equal to that of the corresponding \lcdm\ reference model. 

\begin{figure}
\centering
\includegraphics[width=\linewidth]{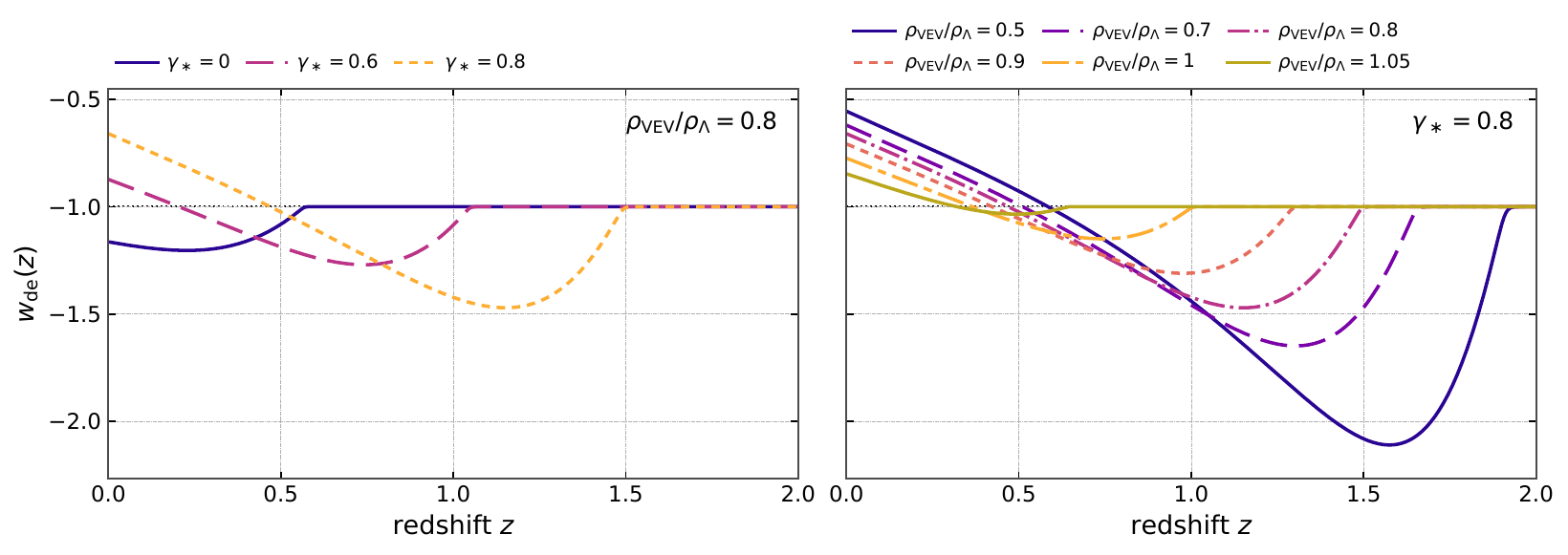}
\caption{
\textit{Left panel:} evolution of $\wde(z)$ for different values of
$\gamma_*=\Gamma/H(z_*)=0,0.6,0.8$, fixing $\rvev/\rhol=0.8$.
\textit{Right panel:} evolution of $\wde(z)$ for different values of
$\rvev/\rhol=0.5,0.7,0.8,0.9,1,1.05$, fixing $\gamma_*=0.8$.
Increasing $\rvev/\rhol$ tends to lower $z_*$, leaving less time
for the decay to affect the evolution of $\wde(z)$.
}
\label{fig:wzcartoon}
\end{figure}

\Cref{fig:wzcartoon}
shows the effective
dark energy equation of state, $w_{\rm de}(z)$, and the effect on it
of the two free parameters of the theory, the scalar field
VEV $\rvev$ before the phase transition and the decay rate
$\Gamma$ after the phase transition.
The equation of state goes phantom immediately after the
phase transition and, in the absence of decay, gradually
heads toward a de Sitter future ($w=-1$). The decay 
(creating relativistic particles with $w_{vd}=1/3$)
instead uplifts the late-time evolution across the
phantom divide $w=-1$.
In the future the scalar field would entirely decay,
leaving $\wde=1/3$ (or eventually $\wde=0$ if the light
particles become nonrelativistic).

In the left panel, increasing $\gamma_\ast$
at fixed $\rvev/\rhol=0.8$ raises the present-day
equation of state.
For $\gamma_\ast=0$, i.e.\ without decay, $w_{\rm de}$
remains phantom throughout the post-transition evolution,
whereas for $\gamma_\ast=0.6$ and $0.8$ it crosses
$w=-1$ at $z\simeq0.20$ and $0.47$, reaching
$w_{{\rm de},0}\simeq-0.87$ and $-0.66$, respectively.

The right panel shows that increasing
$\rvev/\rhol$ at fixed $\gamma_\ast=0.8$ 
lengthens the \lcdm-like high-redshift history,
reducing the time available for decays,
and so keeps $w_0$ closer to $-1$ after crossing the phantom
divide. 
As $\rvev/\rhol$ increases from $0.5$ to $1$, the
transition redshift decreases from $z_*\simeq1.91$ to $1.00$,
while $w_0$ changes from $-0.56$ to $-0.77$. 
Of course, when both parameters vary simultaneously, one 
can get a trade-off in effects.

\section{Cosmological Data and Numerical Code} \label{sec:datacode} 

We implement the background evolution and linear perturbations in a modified version of \texttt{CLASS}~\cite{Lesgourgues:2011re,Blas:2011rf}.
For numerical stability, we regularize the transition using
the hyperbolic tangent window
$S(N)=\{1+\tanh[(N-N_*)/\Delta]\}/2$, where $N=\ln a$ and $N_*=-\ln(1+z_*)$. The transition is implemented through source terms proportional to $\mathrm{d}S/\mathrm{d}N$, chosen to reproduce the matching conditions in Eqs.~(\ref{eq:vli})--(\ref{eq:vdi}) while preserving total energy conservation.
The subsequent decay is activated through $S(N)$, and the instantaneous-transition limit is recovered as $\Delta\rightarrow0$. Throughout our analysis, we fix $\Delta=5\cdot10^{-3}$.\footnote{We have verified that this choice produces completely negligible differences from the instantaneous-transition background at the precision of the data considered, while ensuring numerical stability during the transition.}

For linear perturbations, we work in synchronous gauge. To avoid divisions by vanishing densities, we evolve density and momentum perturbations rather than fractional density contrasts.
The perturbation equations and corresponding pressure relations are presented in Appendix~\ref{sec:apxpert}.
Standard matter and radiation keep their usual \texttt{CLASS} evolution.

We test the model against three data combinations: BAO+SN, BAO+CMB, and BAO+CMB+SN.
Here, BAO denotes the DESI DR2 distance measurements from galaxies, quasars, and the Ly$\alpha$ forest~\cite{DESI:2025qqy,DESI:2025zpo}, including their full covariance~\cite{DESI:2025zgx}.
SN denotes the DES-Dovekie Type Ia supernova sample~\cite{DES:2025sig}, used without an external absolute-magnitude calibration.
The CMB combination includes temperature and polarization measurements from Planck, ACT DR6, and SPT-3G, together with CMB lensing.
We use the Planck 2018 low-$\ell$ TT \texttt{Commander} likelihood~\cite{Planck:2019nip}, the low-$\ell$ EE \texttt{SRoll2} likelihood~\cite{Delouis:2019bub,Pagano:2019tci}, 
and the Planck high-$\ell$ temperature and polarization likelihood~\cite{Planck:2019nip}, restricted to $\ell_{\max}^{TT}=1000$ and $\ell_{\max}^{TE}=\ell_{\max}^{EE}=600$.
ACT DR6 TT, TE, and EE measurements are included over $600\leq\ell\lesssim6300$~\cite{AtacamaCosmologyTelescope:2025blo}, while SPT-3G temperature and polarization measurements are included through the D1 lite likelihood~\cite{SPT-3G:2025bzu}.
For CMB lensing, we include the joint ACT+Planck baseline likelihood~\cite{Carron:2022eyg,ACT:2023dou,ACT:2023kun} and the SPT-3G \texttt{MUSE} lensing likelihood~\cite{SPT-3G:2024atg}.

We explore the parameter space of the model by performing Markov chain Monte Carlo (MCMC) sampling with \texttt{Cobaya}~\cite{Torrado:2020dgo} and analyze the resulting chains with \texttt{GetDist}~\cite{Lewis:2019xzd}. 
In the MCMC, we vary the six standard \lcdm\ parameters\footnote{For BAO+SN we fix $\tau$, $\ln(10^{10}A_s)$, and $n_s$.} together with two additional VM parameters, which we choose for numerical convenience as\footnote{Sampling directly in $\gamma_*$ and
$\rvev/\rho_\Lambda$ can make the closure residual
non-monotonic in $z_*$.
This complicates bracketing and root selection,
making the shooting less robust numerically.}
\begin{equation}
\gamma_0\equiv\frac{\Gamma}{H_0}\ ,
\qquad
\fvev\equiv\frac{\rvev}{\rho_m(z_*)}\ .
\end{equation}
The transition redshift $z_*$ is determined by imposing the present-day closure condition $H(a=1)=H_0$.\footnote{When sampling the acoustic scale $100\,\theta_s$, $H_0$ is determined simultaneously by matching its sampled value.} The physical quantities introduced in \cref{sec:implementation} are then derived for each sample as
\begin{equation}
\gamma_*=\gamma_0\,\frac{H_0}{H(z_*)}\ ,
\qquad
\frac{\rvev}{\rho_\Lambda}
=\fvev\,\frac{\rho_m(z_*)}{\rho_\Lambda}\ ,
\end{equation}
where $\rho_\Lambda$ denotes the \lcdm\ reference density computed using the same $H_0$ and standard-matter $\Omega_m$ as the corresponding VM sample. We always assume one massive and two massless neutrino species with 
$\sum m_\nu=0.06\,{\rm eV}$ and $N_{\rm eff}=3.044$.
The priors adopted for all sampled parameters are listed in Table~\ref{tab:priors}.

\begin{table*}[t]
\centering
\begin{tabular}{l@{\hspace{0.5cm}}c@{\hspace{0.5cm}}c@{\hspace{0.5cm}}c}
\hline
Parameter & BAO+SN & BAO+CMB & BAO+CMB+SN \\
\hline
$\omega_b$ & $\mathcal{N}(0.0224,0.0005)$
& $[0.005,0.1]$ & $[0.005,0.1]$ \\
$\omega_{\rm cdm}$ & $[0.001,0.99]$ & $[0.001,0.99]$ & $[0.001,0.99]$ \\
$100\theta_s$ & $[0.5,10]$ & $[0.5,10]$ & $[0.5,10]$ \\
$\tau$ & $0.055$ & $[0.01,0.8]$ & $[0.01,0.8]$ \\
$\ln(10^{10}A_s)$ & $3.044$ & $[1.61,3.91]$ & $[1.61,3.91]$ \\
$n_s$ & $0.966$ & $[0.8,1.2]$ & $[0.8,1.2]$ \\
$\gamma_0$ & $[0,4]$ & $[0,4]$ & $[0,4]$ \\
$\fvev$ & $[0,3]$ & $[0,3]$ & $[0,3]$ \\
\hline
\end{tabular}
\caption{Priors and fixed values used in the VM analysis.
Intervals denote uniform priors, while
$\mathcal{N}(\mu,\sigma)$ denotes a Gaussian prior
with mean $\mu$ and standard deviation $\sigma$.}
\label{tab:priors}
\end{table*}

Finally, we compare the VM model with \lcdm\ and
$\wowa$CDM using the same data combinations and the same priors
on the common cosmological parameters. For $\wowa$CDM, we adopt uniform priors $w_0\in[-3,1]$ and $w_a\in[-3,2]$. The comparison includes both marginalized MCMC constraints and best-fit predictions. The best-fit parameters and minimum $\chi^2$ values are obtained from dedicated minimizations using the \texttt{BOBYQA} minimizer~\cite{BOBYQA} for each model and data combination.

\section{Results}\label{sec:results}

The parameter constraints are shown in \cref{tab:vm_constraints}. 
For each dataset combination, the standard primordial parameters 
and late-time parameters such as $H_0$ and $\Om$ agree well 
with those from a \lcdm\ or $\wowa$CDM model. The BAO+CMB and 
the full combination BAO+CMB+SN show a clear preference for 
the phase transition, with $z_*$ distinct from zero (which 
would give simply \lcdm).

\begin{table}[htpb!]
\centering\small
\setlength{\tabcolsep}{8pt}
\renewcommand{\arraystretch}{1.08}
\begin{tabular}{lccc}
\toprule
Parameter & BAO + SN & BAO + CMB & BAO + CMB + SN \\
\midrule
$\omega_b$
& $0.02239\pm0.00052$
& $0.022435\pm0.000091$
& $0.022468\pm0.000093$ \\[-2pt]
& {\scriptsize$[0.022756]$}
& {\scriptsize$[0.0224484]$}
& {\scriptsize$[0.0224154]$} \\[3pt]

$\omega_{\rm cdm}$
& $0.1170\pm0.0051$
& $0.11905\pm0.00091$
& $0.11798\pm0.00072$ \\[-2pt]
& {\scriptsize$[0.11726]$}
& {\scriptsize$[0.119216]$}
& {\scriptsize$[0.118673]$} \\[3pt]

$100\theta_s$
& $1.0424\pm0.0062$
& $1.04171\pm0.00023$
& $1.04180\pm0.00022$ \\[-2pt]
& {\scriptsize$[1.04277]$}
& {\scriptsize$[1.041709]$}
& {\scriptsize$[1.041751]$} \\[3pt]

$\ln(10^{10}A_s)$
& $3.044$ (fixed)
& $3.058\pm0.012$
& $3.068\pm0.011$ \\[-2pt]
&
& {\scriptsize$[3.0558]$}
& {\scriptsize$[3.0578]$} \\[3pt]

$n_s$
& $0.966$ (fixed)
& $0.9711\pm0.0033$
& $0.9736\pm0.0031$ \\[-2pt]
&
& {\scriptsize$[0.97118]$}
& {\scriptsize$[0.97198]$} \\[3pt]

$\tau$
& $0.055$ (fixed)
& $0.0615\pm0.0064$
& $0.0668\pm0.0064$ \\[-2pt]
&
& {\scriptsize$[0.06085]$}
& {\scriptsize$[0.05889]$} \\[3pt]
\midrule

$\gamma_0=\Gamma/H_0$
& $2.21\pm0.91$
& $2.24\pm0.52$
& $2.09\pm0.93$ \\[-2pt]
& {\scriptsize$[4.000]$}
& {\scriptsize$[2.385]$}
& {\scriptsize$[1.413]$} \\[3pt]

$f_{\rm VEV}=\rho_{\rm VEV}/\rho_m(z_\ast)$
& $1.32\pm0.38$
& $0.27\pm0.48$
& $1.21\pm0.54$ \\[-2pt]
& {\scriptsize$[1.619]$}
& {\scriptsize$[0.086]$}
& {\scriptsize$[0.215]$} \\[3pt]
\midrule
\midrule

$z_\ast$
& $0.25\pm0.17$
& $1.38\pm0.48$
& $0.35\pm0.35$ \\[-2pt]
& {\scriptsize$[0.140]$}
& {\scriptsize$[1.604]$}
& {\scriptsize$[1.100]$} \\[3pt]

$\gamma_\ast=\Gamma/H(z_\ast)$
& $1.95\pm0.89$
& $1.00\pm0.43$
& $1.80\pm0.96$ \\[-2pt]
& {\scriptsize$[3.635]$}
& {\scriptsize$[0.925]$}
& {\scriptsize$[0.752]$} \\[3pt]

$\rho_{\rm VEV}/\rho_\Lambda$
& $1.072\pm0.039$
& $0.84\pm0.18$
& $1.033\pm0.072$ \\[-2pt]
& {\scriptsize$[1.0878]$}
& {\scriptsize$[0.805]$}
& {\scriptsize$[0.9134]$} \\[3pt]
\midrule

$H_0$ [km s$^{-1}$ Mpc$^{-1}$]
& $67.24\pm0.92$
& $64.7\pm1.5$
& $67.08\pm0.57$ \\[-2pt]
& {\scriptsize$[67.141]$}
& {\scriptsize$[64.06]$}
& {\scriptsize$[67.183]$} \\[3pt]

$\Omega_m$
& $0.3096\pm0.0079$
& $0.340\pm0.016$
& $0.3136\pm0.0057$ \\[-2pt]
& {\scriptsize$[0.31202]$}
& {\scriptsize$[0.3468]$}
& {\scriptsize$[0.31401]$} \\[3pt]

$S_8=\sigma_8\sqrt{\Omega_m/0.3}$
& ---
& $0.842\pm0.011$
& $0.8251\pm0.0069$ \\[-2pt]
&
& {\scriptsize$[0.8456]$}
& {\scriptsize$[0.82876]$} \\[3pt]

$r_d$ [Mpc]
& $147.9\pm1.6$
& $147.28\pm0.24$
& $147.53\pm0.20$ \\[-2pt]
& {\scriptsize$[147.40]$}
& {\scriptsize$[147.218]$}
& {\scriptsize$[147.399]$} \\[3pt]
\midrule
\midrule

$\chi^2_{\rm total}$ & $1637.50$ & $999.28$ & $2637.61$ \\[1pt]
DESI DR2 BAO & $9.82$ & $6.76$ & $13.17$ \\[1pt]
DES-Dovekie SN & $1627.68$ & --- & $1630.15$ \\[1pt]
Planck low-$\ell$ TT & --- & $21.84$ & $21.79$ \\[1pt]
Planck low-$\ell$ EE (Sroll2) & --- & $390.67$ & $390.27$ \\[1pt]
Planck high-$\ell$ (ACT cut) & --- & $220.70$ & $220.04$ \\[1pt]
ACT DR6 (TT/TE/EE) & --- & $157.52$ & $159.37$ \\[1pt]
ACT+Planck lensing & --- & $19.68$ & $20.14$ \\[1pt]
SPT-3G D1 (TT/TE/EE) & --- & $161.35$ & $161.23$ \\[1pt]
SPT-3G MUSE & --- & $20.76$ & $21.45$ \\[1pt]
\midrule
$\Delta\chi^2_{{\rm VM}-\Lambda{\rm CDM}}$
& $-7.82$ & $-12.13$ & $-8.54$ \\[1pt]
$\Delta\chi^2_{{\rm VM}-w_0w_a{\rm CDM}}$
& $-0.62$ & $-0.25$ & $+5.27$ \\[1pt]
\bottomrule
\end{tabular}
\caption{
Constraints on the cosmological and model parameters
for different dataset combinations.
We report posterior means and standard deviations,
with bestfit values from dedicated $\chi^2$ minimizations,
including the profile-likelihood scan, given in square brackets.
Parameters above the first double rule are sampled
(or fixed, where indicated), while those between the
double rules are derived.
The minimum data $\chi^2$, its breakdown by dataset,
and the differences relative to $\Lambda$CDM and $w_0w_a$CDM
are given at the bottom.
}
\label{tab:vm_constraints}
\end{table} 

\begin{figure}[htpb!]
\centering
\includegraphics[width=0.65\linewidth]{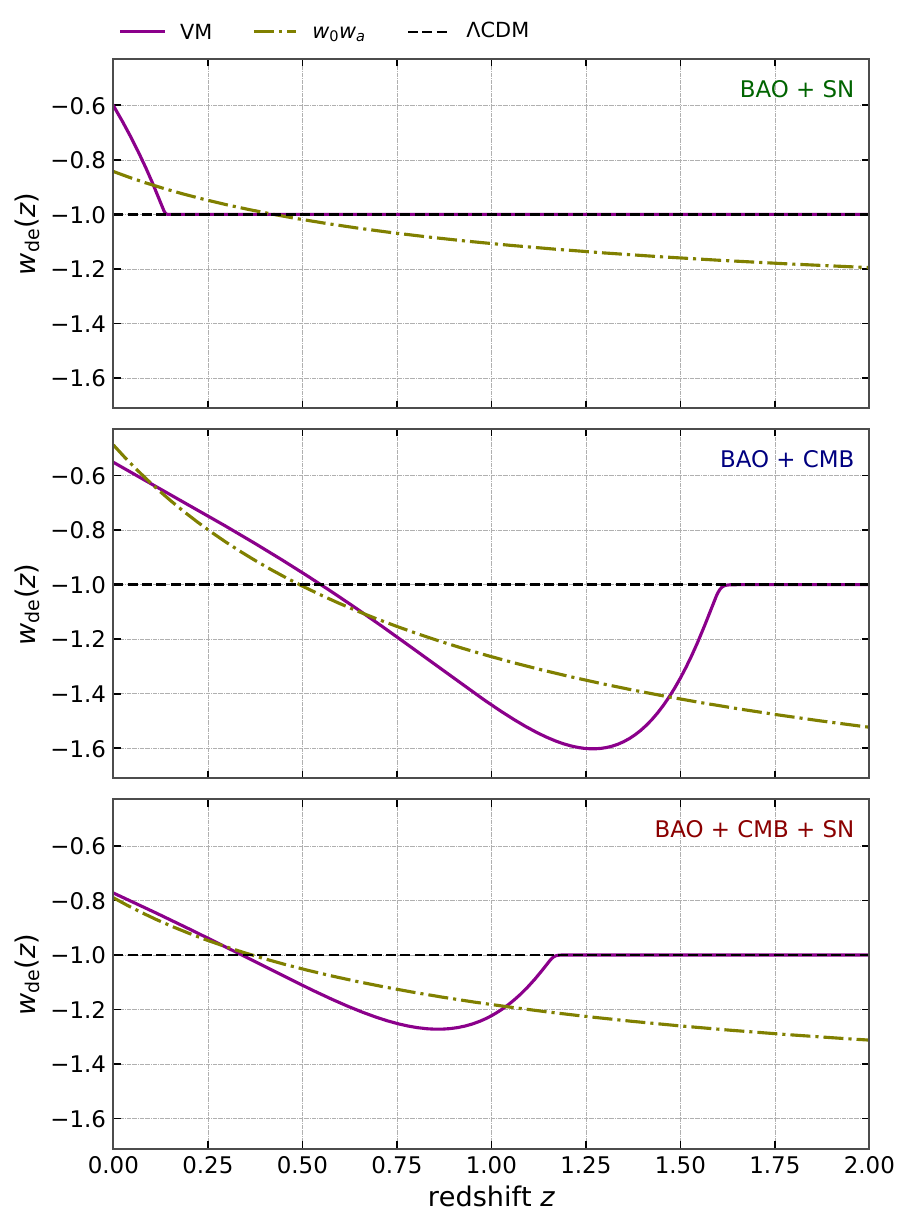}
\caption{The bestfit $w_{\rm de}(z)$ in the VM model (solid curve) for different dataset combinations. The dot-dash curve shows the result for $\wowa$CDM. We emphasize that $w=-1$ earlier than the phase transition does not mean standard \lcdm, as $\rvev$ can be different from the $\rhol$ determined 
by $\Om$.
}
\label{fig:wzbestfit}
\end{figure}

\begin{figure}[htpb!]
    \centering
   \includegraphics[width=\linewidth]{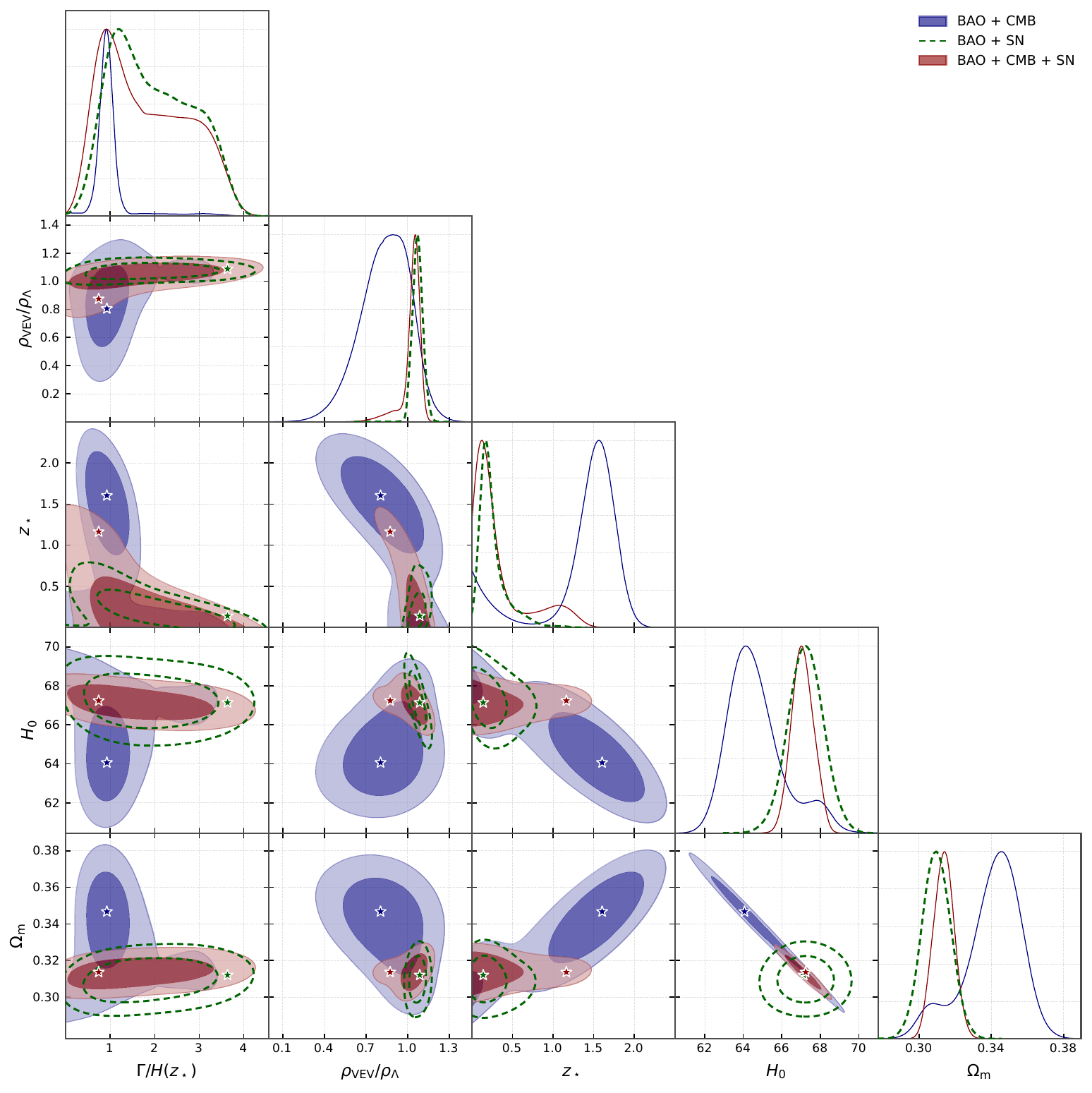}
    \caption{Triangle plot of posterior probability contours, 
at 68\% and 95\% CL, of key cosmological parameters plotted 
for the different combinations of datasets. Some of the 
posteriors are noticeably non-Gaussian, and the bestfit points 
are indicated by stars. 
The bestfits can lie outside the 68\% CL 
contour but are within the 95\% CL contour; this is mainly due 
to the parameter volume effect discussed in the text.
}
\label{fig:tplotall}
\end{figure}

The BAO+SN combination accepts a 
low $z_*$, but with an important difference from \lcdm: the 
cosmological constant (VEV) at $z>z_*$ is not the value $\rho_\Lambda$ 
expected from $\Omega_\Lambda=1-\Om$, but rather 
$\rvev/\rho_\Lambda=1.09$. For this combination, the acceleration 
is strengthened over \lcdm\ at higher redshifts not by having 
phantom dark energy, but by increasing $\rvev$ above $\rho_\Lambda$. This will not fit CMB data well since it 
shifts the distance to last scattering, as opposed to the 
phantom-crossing dark energy that preserves the distance to 
last scattering by compensating the high acceleration at 
higher redshifts (from $w<-1$) with a lesser acceleration at 
lower redshifts (from $w>-1$). Thus, we do not consider such a 
low-redshift phase transition as a case of overall interest 
(despite its $\chi^2$ being better than \lcdm\ by $-7.8$) -- 
the CMB is an important dataset to fit!

For the BAO+CMB combination, the best-fit value is $z_*=1.6$. Since such 
a high $z_*$ gives a strongly phantom $w_{\rm de,min}=-1.6$ after the phase 
transition, to preserve the distance to last scattering this 
case actually prefers $\rvev<\rho_\Lambda$. The decay rate 
$\Gamma/H(z_*)$ is close to the natural value of unity. The 
minimum $\chi^2$ is lower than that of \lcdm\ by $12.1$, indicating an 
improved fit for vacuum metamorphosis, and is comparable 
to that of $\wowa$CDM, lower by $0.25$ for the same number of free parameters.

With the full data combination, the bestfits are mostly in 
between the two previous cases. However, the posteriors are 
highly non-Gaussian, with the bestfit offset from the posterior 
mean (though within the 95\% CL contour, as we will see). The bestfit value is $z_*=1.1$, with
$w_{\rm de,min}\simeq-1.2$ and $\rvev\approx0.91\rhol$.
Again, the decay rate is close to the natural value of unity.
The minimum $\chi^2$ for vacuum metamorphosis
is lower than that of \lcdm\ by $8.5$, but is $5.3$ higher than
that of $\wowa$CDM. About half of this comes from BAO,
$0.4$ from SN, and about $1.9$ from CMB,
with $1.4$ arising from ACT DR6.
Still, vacuum metamorphosis with decay appears to give a 
reasonable fit to the data, along with a physical basis. 

Figure~\ref{fig:wzbestfit} shows the bestfit dark energy equation of state $\wde(z)$ for 
various dataset combinations and compares it to that from 
$\wowa$CDM. Before the phase transition, $z>z_*$, the VEV 
gives $\wde=-1$; then, after the phase transition, $z<z_*$, the 
behavior becomes phantom and gradually bends upward and crosses 
$w=-1$ to reach $w_0>-1$ due to the scalar field decays. 

The mean parameter values for the dark energy parameters 
are marginalized over the full parameter space and sometimes 
are offset from the bestfit parameter values. 
Figure~\ref{fig:tplotall} shows the triangle plot of 
posterior probabilities with 68\% and 95\% CL, along with 
stars denoting the bestfit values. Two issues interact to 
give a prior volume effect enhancing the non-Gaussian aspect 
of the contours. At low $z_*$, the value of the decay rate 
$\Gamma/H(z_*)$ is mostly moot, as there is insufficient time 
after the phase transition for the scalar field to decay 
significantly. In addition, at $z_*\lesssim0.5$ the dark energy equation of state 
$\wde(z)$ is always close to $-1$, and so the exact value of 
$z_*$ is mostly unimportant, as was seen in 
Figure~\ref{fig:wzcartoon}. Overall, 
regions with small $z_*$ can cover a large volume of parameter space throughout which observables would look similar to $\Lambda$CDM; while not a bestfit, the large volume can 
weigh into the marginalization to pull the mean parameter 
values toward this region. 
See Appendix~\ref{sec:apxsuppl} for more details.

\clearpage
\section{Discussion and Conclusions}\label{sec:concl} 

The vacuum metamorphosis scenario provides a concrete physical 
mechanism for generating an effective dark energy sector whose 
equation of state can cross the phantom divide. The phase 
transition in the scalar field occurs when the Ricci scalar 
reaches a critical value related to the scalar field mass, 
driven by the fully resummed quantum gravity loop corrections 
calculated by Parker and colleagues~\cite{Parker:1999fc,Parker:2001ws,Parker:2003as,Caldwell:2005xb}. After the transition, the 
Ricci scalar freezes through a Lenz's law-type feedback, causing 
the scalar field to act phantom, $\wde<-1$. Decay of the scalar 
field into light particles after the transition can then 
eventually uplift the scalar field to cross the phantom divide. 

The model introduces two additional late-time parameters relative 
to \lcdm: the decay rate $\gamma_*=\Gamma/H(z_*)$ and the VEV, 
which can be written in terms of the vacuum 
energy fraction $\rvev/\rhol$. Thus, the model matches the 
parameter count of $w_0w_a$CDM. We implemented this model in a modified Boltzmann solver and constrained it with DESI DR2 BAO, current CMB data, and a recent SNe~Ia distance compilation.

The phantom divide is successfully crossed at $z\approx0.5$ 
in this physics-based 
theory, with a natural value of $\gamma_*\approx1$. Comparison 
with data shows overall improvement by $\Delta\chi^2=-8.5$ ($-12$) 
relative to \lcdm\ for BAO+CMB+SN (BAO+CMB). With respect to 
$\wowa$CDM, $\Delta\chi^2=+5$ ($-0.2$). As to the origin of the $\chi^2$ difference, at the joint BAO+CMB+SN bestfit, the stronger phantom phase in VM reduces the dark energy density at intermediate redshifts $0.5\lesssim z \lesssim 1$ relative to $\wowa$CDM, lowering $H(z)$ and increasing the radial BAO distance $D_H/r_d$. 
One can get an intuitive sense for this by noticing 
that in Fig.~\ref{fig:wzbestfit} the bestfit 
$w(z)$ curve for VM in the BAO+CMB+SN case lies below or even with the curve 
for $\wowa$CDM, while in the BAO+CMB case (where VM 
does as well in $\chi^2$ as $\wowa$CDM) the $w(z)$ 
curves have compensating above and below deviations, 
especially out to $z\approx1$. 
The increased $D_H/r_d$ for VM in the BAO+CMB+SN case worsens the agreement particularly with the radial BAO measurements at $z=0.706$ and $0.934$. 
The smaller transverse distance at $z=0.51$ contributes to the BAO penalty as well. The different expansion history also introduces small shifts in the preferred physical densities. In particular, we obtain a lower $\omega_{\rm cdm}$ compared to $\wowa$CDM, which delays matter-radiation equality and modifies the decay of gravitational potentials around horizon entry, changing the relative peak heights in the CMB and contributing to the modest increase in $\chi^2$ for the ACT spectra. 
These are individually small effects but sufficient collectively 
to increase the $\chi^2$ by $\sim 5$.

Overall, vacuum metamorphosis with decay can provide a cogent 
physical explanation of the cosmology preferred by current data, with 
a stronger accelerating component in the past, phantom crossing 
at $z\approx0.5$ and $w_0\approx-0.8$. Future work will include 
growth of structure data (though the $S_8$ value given by 
\texttt{CLASS} agrees well with the $\wowa$CDM value). A promising 
avenue involves surveys capable of further detailing the 
expansion history at $z\gtrsim1.5$, to constrain more tightly 
$z_*$ and $\rvev/\rhol$.

\acknowledgments
The authors acknowledge the use of computational resources provided by the high-performance computing clusters Seondeok at KASI, Stanage at the University of Sheffield, and KOA at the University of Hawai‘i.
EDV is supported by a Royal Society Dorothy Hodgkin Research Fellowship. This article is based upon work from the COST Action CA21136 - ``Addressing observational tensions in cosmology with systematics and fundamental physics (CosmoVerse)'', supported by COST - ``European Cooperation in Science and Technology''. 
EL thanks the Simons Foundation for support as a
Simons Visiting Scientist at the Galileo Galilei Institute, where some of this work took place. WG~acknowledges support from National Aeronautics and Space Administration (NASA) under Grant No.~80NSSC24K0898. KL acknowledges support from the U.S. Department of Energy under Grant No. DE-SC0016021.


\bibliographystyle{apsrev4-1}
\bibliography{newbiblio}


\appendix

\clearpage

\section{Perturbation Equations} \label{sec:apxpert} 

To describe the linear perturbations, we start from a covariant expression of the conservation and flux between species,
\begin{eqnarray}
    \nabla_\mu T^{\mu\nu}_{v\lambda} &=& \Gamma u_{\mu}^{v\lambda} T^{\mu\nu}_{v\lambda} \\
    \nabla_\mu T^{\mu\nu}_{vm} &=& \Gamma u_{\mu}^{vm} T^{\mu\nu}_{vm}\\
    \nabla_\mu T^{\mu\nu}_{vr} &=&  \Gamma u_{\mu}^{vr} T^{\mu\nu}_{vr}\\
    \nabla_\mu T^{\mu\nu}_{vd} &=& -\Gamma \left(u_{\mu}^{v\lambda} T^{\mu\nu}_{v\lambda}+ u_{\mu}^{vm}T^{\mu\nu}_{vm} + u_{\mu}^{vr}T^{\mu\nu}_{vr} \right).
\end{eqnarray}
For our purposes, each stress-energy tensor is modeled as an ideal fluid at the background level, whereby $T^{\mu\nu} = (\rho + p) u^\mu u^\nu + p g^{\mu\nu}$ for each component, and $u^\mu$ is the 4-velocity of the relevant fluid. For linear perturbations, we follow the notation of Ma \& Bertschinger \cite{Ma:1995ey}, and perturb $\rho \to \rho + \delta\rho$, $p \to p+\delta p$. In the conformal Newtonian gauge, each fluid four-velocity is $u^\mu = (1-\Psi, v^i)/a$; in the synchronous gauge, $u^\mu = (1, v^i)/a$. Furthermore, we use a gauge-invariant definition of the pressure perturbation
\begin{equation}
    \delta p_i = c_{s,i}^2 \delta\rho_i + 3 {\cal H} (c_{s,i}^2 - c_{a,i}^2)(\rho_i + p_i)\theta_i/k^2,
\end{equation}
where $c_{a,i}^2 = p_i^\prime / \rho_i^\prime$. Recall that those 
components denoted $vm$, $vr$, and $v\lambda$ are merely 
convenient ways of talking about aspects of the single scalar 
field. Therefore, like a standard scalar field, they have 
$c_{s,vm}^2=c_{s,vr}^2=c_{s,v\lambda}^2=1$ (despite their 
background evolutions having different values of $w$). The decay product component $vd$, on the other hand, is composed of relativistic 
particles and has the corresponding $c_{s,vd}^2=1/3$. The $v\lambda$ component has $w_{v\lambda}=-1$ and does not fluctuate.

The synchronous-gauge fluctuation equations are as follows:
\begin{eqnarray}
    \delta p_{vm} &=& \delta\rho_{vm} + (3 {\cal H} + \Gamma a) \rho_{vm} \theta_{vm} / k^2 \\
    \delta p_{vr} &=& \delta\rho_{vr} + \frac{2}{3} (4 {\cal H} + \Gamma a) \rho_{vr} \theta_{vr}/k^2 \\
    \delta p_{vd} &=& \frac{1}{3} \delta\rho_{vd}  
\end{eqnarray}
\begin{eqnarray}
    (\rho_{vm} \theta_{vm})^\prime &=& -(4 {\cal H} + \Gamma a) (\rho_{vm} \theta_{vm}) + k^2 \delta p_{vm} \\
    (\rho_{vr} \theta_{vr})^\prime &=& -(4 {\cal H} + \frac{3}{4}\Gamma a) (\rho_{vr} \theta_{vr}) + \frac{3}{4} k^2 \delta p_{vr} \\
    (\rho_{vd} \theta_{vd})^\prime &=& -4 {\cal H} (\rho_{vd} \theta_{vd}) + \frac{1}{4} k^2 \delta\rho_{vd} + \frac{3}{4} \Gamma a (\rho_{vm} \theta_{vm} + \rho_{vr} \theta_{vr}) 
\end{eqnarray}
\begin{eqnarray}
    \delta\rho_{vm}^\prime &=& -3 {\cal H}(\delta\rho_{vm} + \delta p_{vm}) - \rho_{vm} \theta_{vm} - \frac{1}{2} \rho_{vm} h^\prime - \Gamma a \delta\rho_{vm} \\
    \delta\rho_{vr}^\prime &=& -3 {\cal H}(\delta\rho_{vr} + \delta p_{vr}) - \frac{4}{3}\rho_{vr} \theta_{vr} - \frac{2}{3} \rho_{vr} h^\prime - \Gamma a \delta\rho_{vr} \\
    \delta\rho_{vd}^\prime &=& -4 {\cal H} \delta\rho_{vd} - \frac{4}{3} \rho_{vd}\theta_{vd} - \frac{2}{3} \rho_{vd} h^\prime + \Gamma a(\delta\rho_{vm} + \delta\rho_{vr})\ .
\end{eqnarray}
Use of the combination $\rho\theta$, i.e.\ the momentum, simplifies the system of equations.

\section{Parameter Distributions and Volume Effect} \label{sec:apxsuppl} 

\begin{figure}
    \centering
   \includegraphics[width=\linewidth]{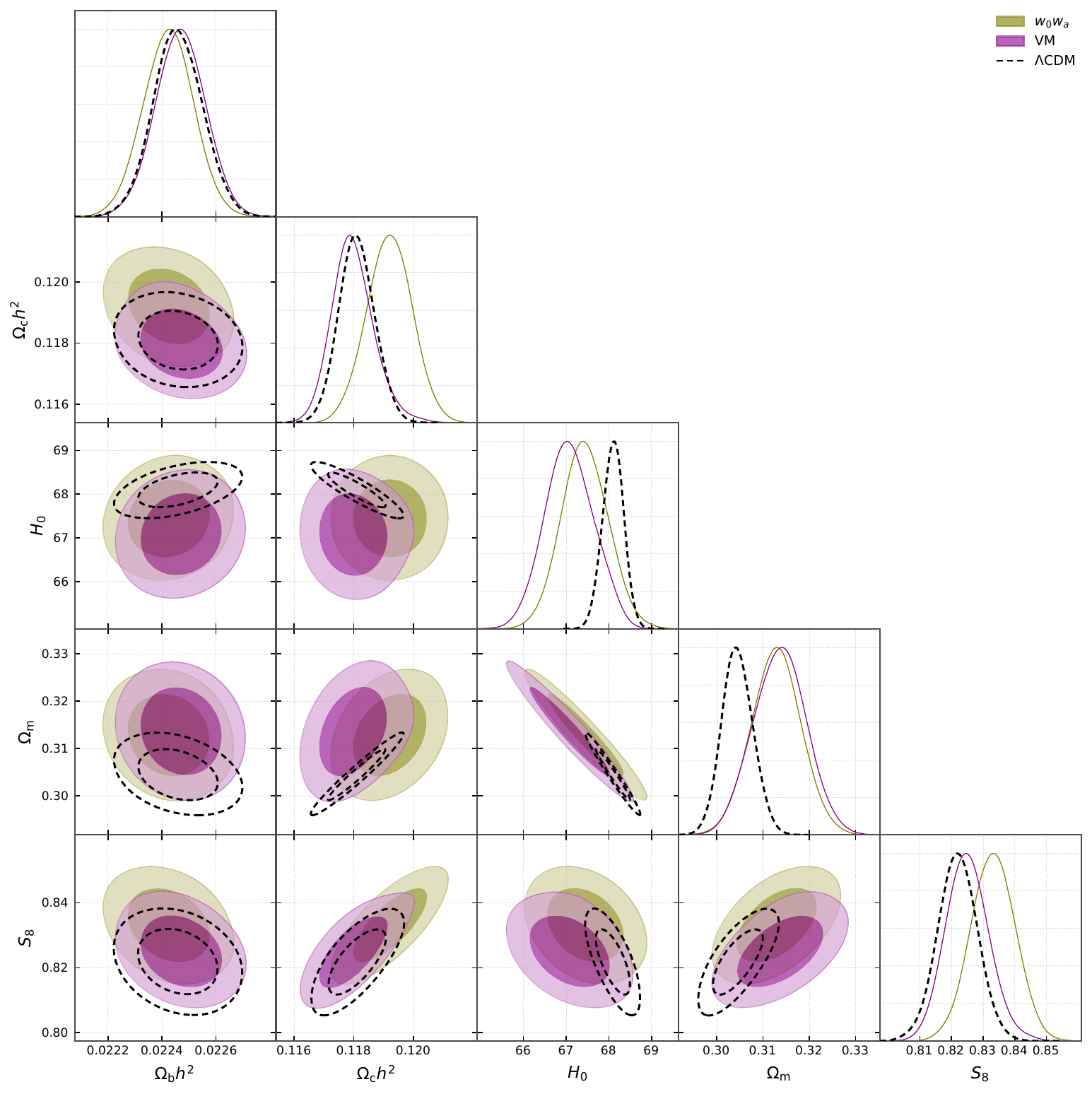}
    \caption{
    Triangle plot of posterior probability contours, 
at 68\% and 95\% CL, of the usual (non-VM model) cosmological parameters plotted 
for the full combination BAO+CMB+SN. The 
posteriors are close to Gaussian.  
    }
    \label{fig:triordpar}
\end{figure}

Cosmological parameters apart from the VM model parameters have 
much the same posterior distributions as in the $\wowa$CDM 
case. Figure~\ref{fig:triordpar} shows the 2D and 1D 
distributions for the usual parameters.

\begin{figure}
    \centering
   \includegraphics[width=0.6\linewidth]{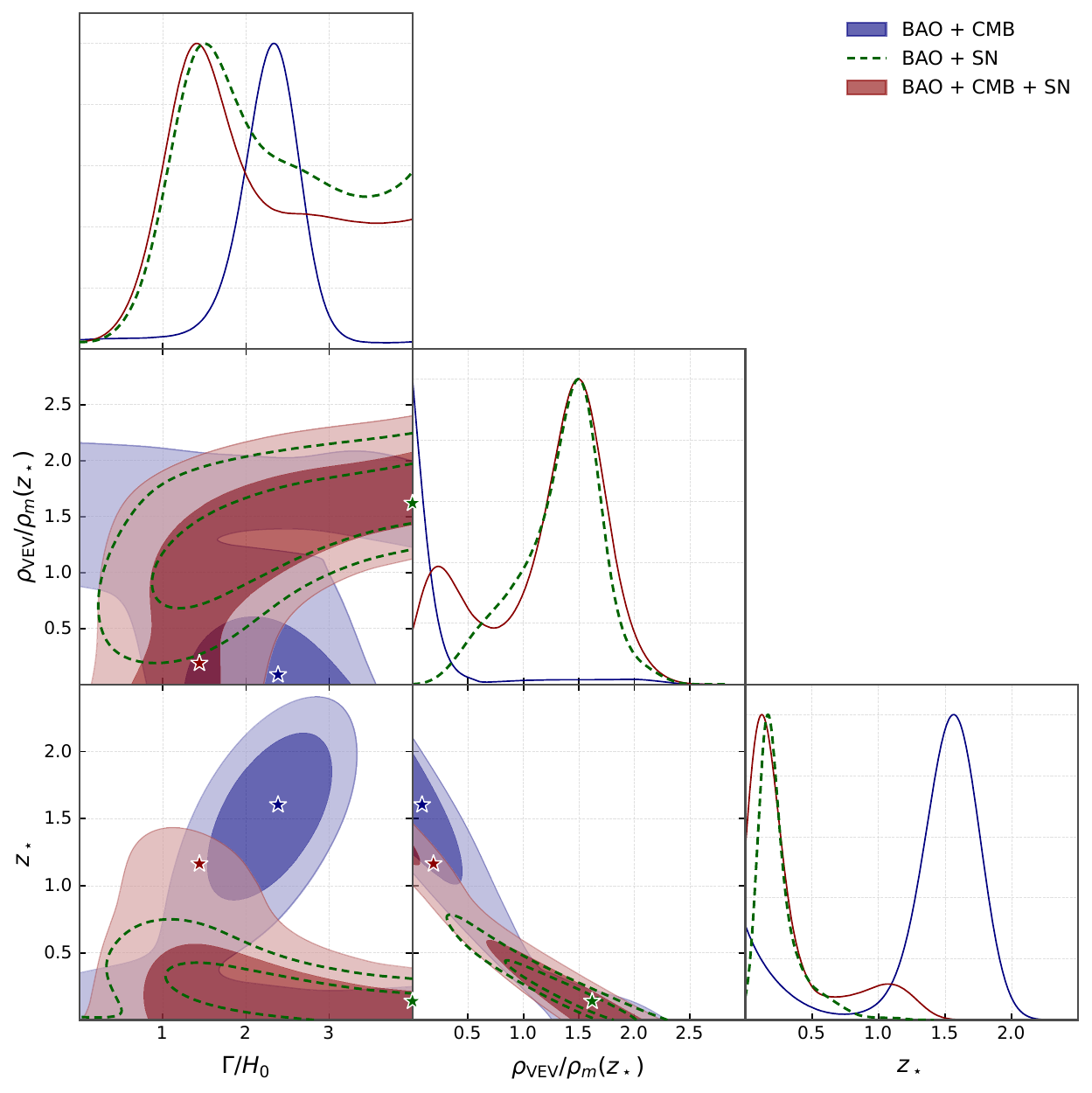}
    \caption{
Triangle plot of posterior probability contours, 
at 68\% and 95\% CL, of the
sampled VM model parameters $\Gamma/H_0$ and
$\rvev/\rho_m(z_*)$, together with the derived transition
redshift $z_*$, for the different dataset combinations. 
Note that at low $z_*$, there is little sensitivity to 
$\Gamma/H_0$ since there is insufficient time for decays to 
have a significant impact. This in turn opens up large values 
of $\rvev/\rho_{m}(z_*)$. Together, these large volumes of 
parameter space with nearly constant $\chi^2$ pull the 
marginalized distributions despite the bestfits (indicated 
by stars) lying at the secondary maxima, as seen in the 1D 
distributions. The parameter volume degeneracy is most 
pronounced for the BAO+CMB+SN full combination. 
    }
    \label{fig:tsampled}
\end{figure}

For an illustration of the parameter volume effect driving some 
of the VM model parameter distributions non-Gaussian, consider 
Figure~\ref{fig:tsampled} for the actual sampled model 
parameters. The distance predictions, and hence the $\chi^2$ 
values of the fits, are nearly the same for late transitions, 
$z_*<0.5$, since the decay rate is mostly moot due to 
insufficient time to have much effect. This spreads the 
posteriors in the sampled parameters $\Gamma/H_0$ (and the 
derived $\Gamma/H(z_*)$ as seen in Fig.~\ref{fig:tplotall}) 
and $\rvev/\rho_m(z_*)$, since lower $z_*$ means lower 
$\rho_m(z_*)$ and hence opens up large $\rvev/\rho_m(z_*)$. 
(In fact, that extra volume closely follows a degeneracy $\rvev/\rhol\approx 1.05$.) These nearly uniform $\chi^2$ 
volumes influence the marginalized distributions, pulling 
them off the bestfits.

\begin{figure}
    \centering
   \includegraphics[width=0.5\linewidth]{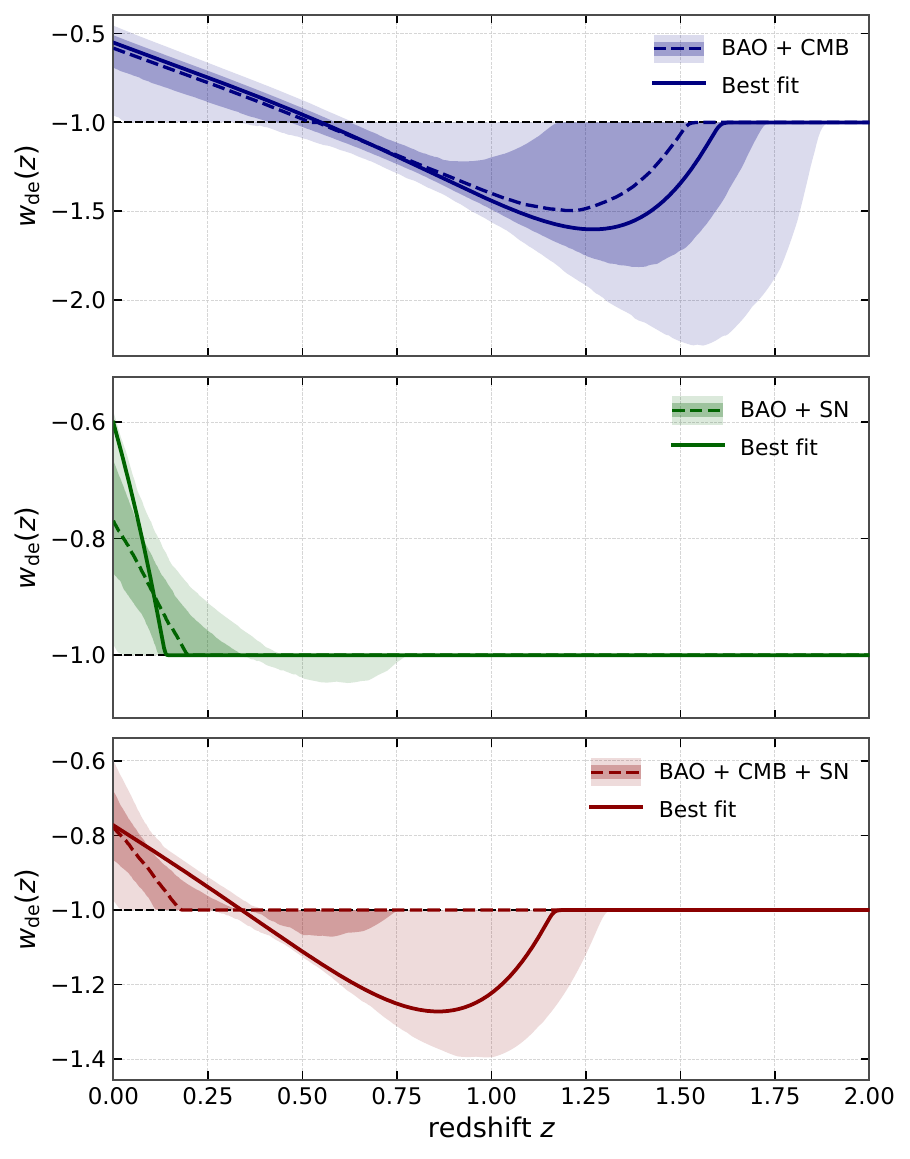}
    \caption{
The $w_{\rm de}(z)$ in the VM decay model for different dataset combinations compared between the bestfit (solid) and 
marginalized (dashed, with 68\% and 95\% CL shaded bands) results. 
The main difference is for the BAO+CMB+SN full combination, 
where the parameter volume effect is worst; the BAO+CMB 
combination actually has good agreement between the bestfit 
and marginalized posteriors. 
Again, $\wde=-1$ earlier than the phase transition does not mean standard 
\lcdm, as $\rvev$ can be different from the $\rhol$ determined 
by $\Om$.
}
\label{fig:wzmargbf}
\end{figure}

\begin{figure}
    \centering
   \includegraphics[width=0.75\linewidth]{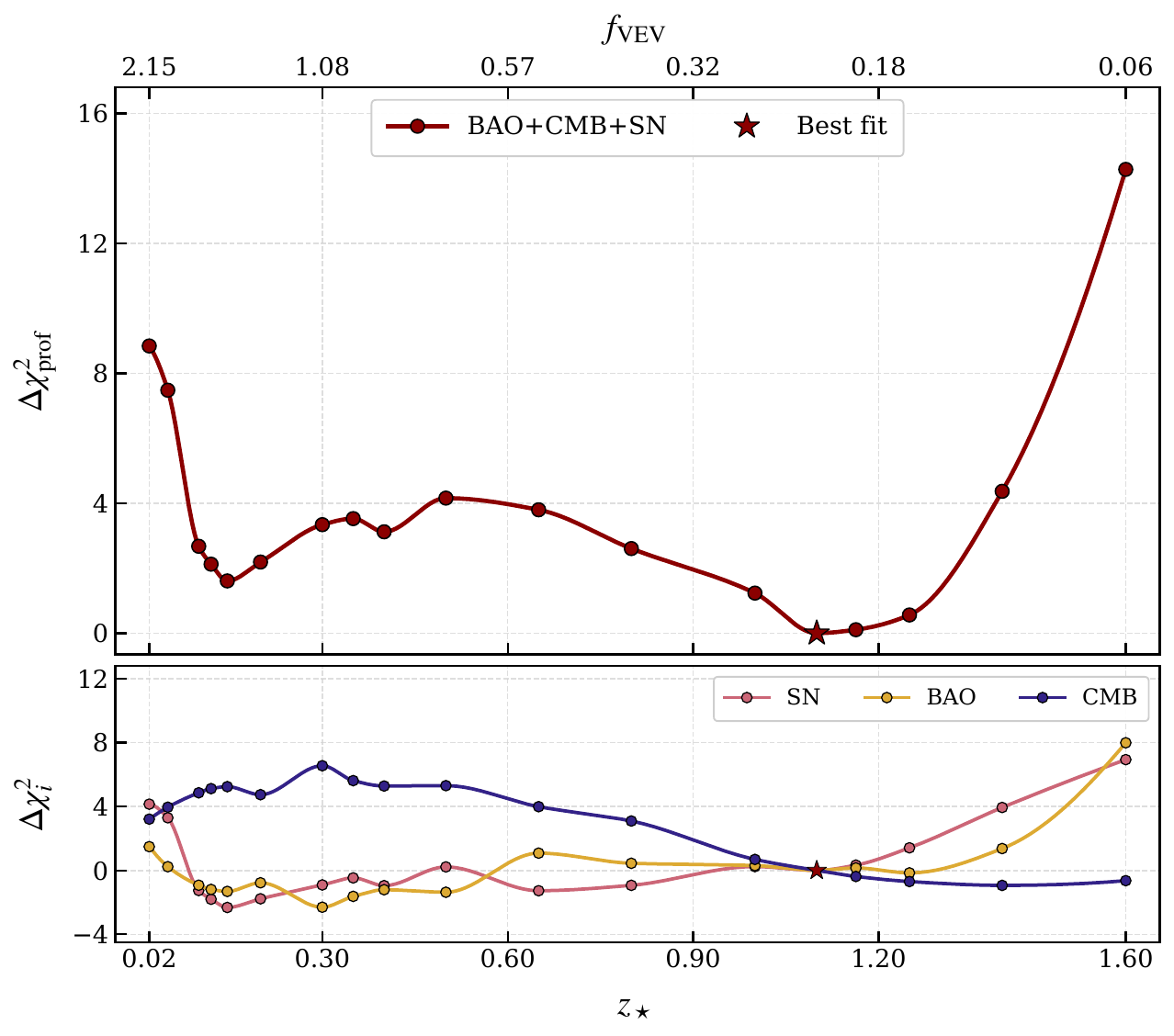}
    \caption{
The profile likelihood for $z_*$ in the BAO+CMB+SN case shows that 
the minimum $\chi^2$ occurs for $z_*\approx1.1$. The secondary 
minimum (worse by $\Delta\chi^2\approx1.6$) is amplified under 
marginalization by the parameter space volume -- e.g.\ for such 
low $z_*$, there is insensitivity to the decay rate. The bottom panel breaks it down by data set.}
    \label{fig:zsprofile}
\end{figure}

The impact of the parameter volume effect is also clearly 
visible in $\wde(z)$. Figure~\ref{fig:wzmargbf} adds to 
the bestfit $\wde(z)$ of Fig.~\ref{fig:wzbestfit} the 
marginalized mean and 68\% and 95\% CL bands for the different 
dataset combinations. Interestingly, the parameter volume 
issue has relatively little effect in the BAO+CMB case, and 
recall that here vacuum metamorphosis gave as good a fit to 
the data as $\wowa$CDM. The major offset of the bestfit from 
the marginalized mean $z_*$ is in the BAO+CMB+SN case. 
Figure~\ref{fig:zsprofile} shows its profile likelihood, and the contributions from each data set, 
demonstrating that indeed the 1D posterior for $z_*$ in 
Fig.~\ref{fig:tsampled} is dominated by the parameter space 
volume effect.

\end{document}